\documentclass{aa}
\usepackage{natbib}
\usepackage{graphicx}
\usepackage{txfonts}

\begin{document}

\title{Intermediate-size fullerenes  as degradation products of interstellar PAHs} 
\titlerunning{Intermediate-size fullerenes}

\author{A. Omont\inst{1}
\and H.\ F.\ Bettinger\inst{2}
}
\institute{}
\institute{
Sorbonne Universit\'e, UPMC Universit\'e Paris 6 and CNRS, UMR 7095, Institut d'Astrophysique de Paris, France\\
\email{omont@iap.fr}
\and  Institut für Organische Chemie, Universität Tübingen, Auf der Morgenstelle 18, 72076 Tübingen, Germany\\
\email{holger.bettinger@uni-tuebingen.de}
}

\abstract{The high interstellar abundances of polycyclic aromatic hydrocarbons (PAHs) and their size distribution are the result of complex chemical processes implying dust, UV radiation, and the main gaseous components (H, C$^+$, and O).  
These processes must explain the high abundance of relatively small PAHs in the diffuse interstellar medium (ISM) and imply the continuous  formation of some PAHs that are small enough (number of carbon atoms N$_{\rm C}$ $<\sim$ 35-50) to be 
completely dehydrogenated by interstellar UV radiation. The carbon clusters C$_{\rm n}$ thus formed are constantly exposed to the absorption of $\sim$10-13.6\,eV UV photons, allowing isomerization and favoring the formation of the most stable  isomers. 
They might tend to form irregular carbon cages. The frequent accretion of 
interstellar C$^+$ ions could favor further cage isomerization, as is known in the laboratory for C$_{60}$, possibly yielding most stable fullerenes, 
such as C$_{40}$, C$_{44}$, and C$_{50}$. 
These fullerenes are expected to be very stable in the diffuse ISM because C$_2$ ejection is not possible by single UV photon absorption, but could need rare two-photon absorption.
It is possible that at least one of these fullerenes or its cation is as abundant as C$_{60}$ or C$_{60}^+$ in the diffuse ISM, although this abundance is limited by the lack of observed  matching  features in observed mid-infrared spectra. 
 B3LYP calculations of the visible spectrum for a number of fullerene isomers  with  40\,$\le$\,N$_{\rm C}$\,$\le$\,50 show that they generally have a few spectral bands in the visible range, with f-values in the range of a few 10$^{-2}$. 
This could make such  fullerenes  interesting candidates for the carriers 
of some diffuse interstellar bands. }


\maketitle 

\section{Introduction}
 
Interstellar carbonaceous macromolecules with a number of carbon atoms,  N$_{\rm C}$ $\sim$ 30-150,  which are intermediate between dust particles and  the  
$\sim$200 small molecules (N$_{\rm C}$ $\la$11) detected to date, mostly by radio astronomy, are an important component of the interstellar medium 
(ISM)  (e.g., Candian et al.\ 2018; Tielens 2013). They are mostly detected through their infrared (IR)  vibrational emission from 3\,$\mu$m to 20\,$\mu$m that results from sporadic IR fluorescence following the absorption 
of a UV photon. They are found in various components of the ISM 
 that are illuminated by UV radiation: In the general ISM of galaxies that is permeated by the UV radiation of massive stars, that is, mostly diffuse interstellar gas and photodissociation regions; and in the more specific environment of young stars and circumstellar matter of evolved stars in the post-asymptotic-giant-branch (post-AGB)
and planetary nebula stages. 

The carbon skeleton of these macromolecules is mostly composed of $sp^2$  
aromatic carbon, either plane polycyclic aromatic hydrocarbons (PAHs) or cage fullerenes (with a more marginal  presence of $sp^3$ micro-diamonds). Interstellar PAHs (L\'eger \& Puget 1985; Allamandola et al.\ 1985, Tielens 2008) are thought to be mostly made of plane fused benzene rings  with 
H-decorated edges, each typically with about 40 to 150 C atoms. Their presence in the ISM is inferred from the strong IR emission in the aromatic bands corresponding to the various CH or C-C vibration modes.  
However, no individual interstellar PAH is identified yet except for CN-naphthalene, which was detected very recently in a dense cloud well shielded from UV radiation (McGuire et al.\ (2021)).  Altogether, PAHs typically contain $\sim$10\% of 
all interstellar carbon, which makes them one of the main reservoirs of interstellar carbon together with C$^+$, CO, and carbon dust. 

Interstellar fullerenes are much less conspicuous  than PAHs. Their typical abundance is a few 10$^{-4}$ of interstellar carbon in various environments (but up to 1\% in special cases of planetary nebulae, where C$_{60}$ was first identified by Cami et al.\ 2010). C$_{60}$ is the only fullerene that is widely detected from its weak mid-IR features (see, e.g., the introduction of Cami et al.\  2018; Bern\'e et al.\ 2015b).  Weaker mid-IR features of 
C$_{60}^+$ are also detected in a few objects (and C$_{70}$ in one object).   C$_{60}^+$ has been widely found  in the diffuse ISM through its strong 
diffuse interstellar bands (DIBs) at 9577\,\AA\ and 9632\,\AA\ (Foing \& Ehrenfreund 1994; Campbell et al.\ 2015). 

The formation of interstellar  C$_{60}$ is still debated (e.g., Candian et al.\ 2018). The most convincing scenarios imply the curvature and closure of sheets of graphene, in which N$_{\rm C}$ is slightly larger than 60 heated by UV photons (see Section 3.1),
or energetic processing of carbon or SiC dust in planetary nebulae (Cami et al.\ 2018; Bernal et al. 2019).
Fullerenes are known to be very stable, especially C$_{60}$.  An energetic process is needed to destroy them, similar to PAHs (see below). However, the lack of CH bonds makes small fullerenes less sensitive to UV photodissociation than PAHs with a similar number of C atoms. 

The identification of DIB carriers 
remains an outstanding problem in astrophysics (see  Cox 2011; Cami \& Cox 2014 and, e.g., Introduction of Omont et al.\ 2019).
 It  is generally agreed that DIB carriers are probably large carbon-based molecules (with $\sim$10-100 atoms) in the gas phase, such as PAHs, fullerenes, or carbon chains. The most stringent constraint on DIB carriers comes from the strength (equivalent width) of the bands, which can be expressed as the product of the carrier abundance by the band oscillator strength  f (e.g., Cami 2014; Omont 2016). The equivalent width per unit reddening may be approximately written as  (for  N$_{\rm C}$\,$\sim$\,60)
\begin{equation}
{\rm EW(mA/mag)   \approx   10^7~f~X_{CM}}
,\end{equation}
where  X$_{{\rm CM}}$ denotes the fraction of total interstellar carbon locked up in the  M 
molecule. 

\begin{figure*}[htbp]
         \begin{center}
\includegraphics[scale=0.75, angle=0]{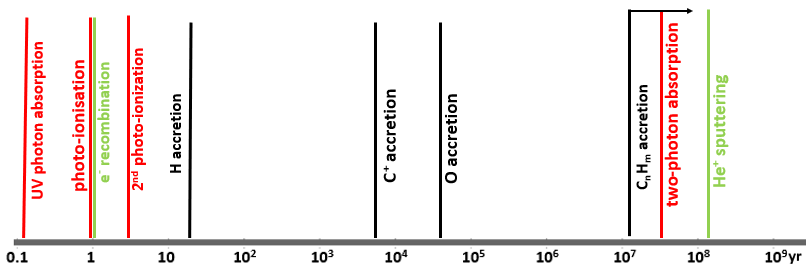}
 \caption{Estimated orders of magnitude of reaction rates for a typical PAH/PAH$^+$ of 50 C atoms in a diffuse interstellar cloud with n$_{\rm H}$\,=\,50\,cm$^{-3}$, T\,=\,50-100\,K, and a UV intensity G$_0$\,=\,3\,Habing (Table A.1).}
     \end{center}
 \end{figure*}

The aim of this paper  is to extend our previous review of the properties of interstellar fullerene compounds (Omont 2016) by pointing out that in addition to  C$_{60}$, other smaller fullerenes with N$_{\rm C}$ between about 40 and 50 could be abundant in the diffuse ISM, resulting from the downsizing, dehydrogenation, and photoprocessing of the smallest interstellar PAHs. 
They could also be interesting candidates for the carriers of DIBs.

This paper is organized as follows. Section 2 and Appendix A summarize a few properties of the smallest interstellar PAHs  in the diffuse ISM that are relevant for their eventual degradation into fullerenes. We focus on their size distribution, chemistry, and evolution. Section 3 argues that in the diffuse ISM, we may expect an efficient formation of carbon cages from the smallest PAHs with N$_{\rm C}$ $\sim$ 40-50 when they are fully dehydrogenated by photodissociation. 
Section 4 addresses the likely evolution of these cages into the most stable fullerenes, such as C$_{50}$ or C$_{44}$, through isomerization, which could be  favored by C$^+$ accretion.
Sputtering of these fullerenes  and the possible resulting abundances are then discussed. 
In Section 5 we finally consider the possible identification of these interstellar  neutral or cation fullerenes through either their mid-IR spectrum or their optical spectrum (for which we performed B3LYP calculations), 
which could make them interesting carriers of DIBs. General conclusions and future prospects are outlined in Section 6.

\section{Size distribution and evolution of the smallest PAHs in the diffuse ISM}

The aim of this section is to recall how we understand the abundance and the size distribution  of the smallest PAHs in the diffuse ISM, before they can eventually transform into fullerenes through downsizing and dehydrogenation. The details of what follows are outlined in Appendix A, which summarizes the  properties of PAHs in the diffuse ISM and their chemical processes and recalls the conclusions of the models of their size distribution.

While the discussions of interstellar PAHs generally refer to regions with strong UV radiation, we here focus on typical interstellar diffuse clouds in which the interstellar absorption leading to DIBs mostly takes place,
with a density n$_{\rm H}$\,=\,30-100\,cm$^{-3}$, temperature T\,=\,50-100\,K, and UV intensity G$_0$\,=\,1-10\,Habing (Table A.1).

The key features of interstellar PAHs have been recognized since their discovery: They are part of a continuous distribution of aromatic carbonaceous particles that extends to the largest dust grains. Small PAHs with N$_{\rm C}$\,$\la$\,30-40 are absent, however, because their CH bonds are photodissociated. Various chemical processes, including reactions with H, C$^+$ and O, should explain their size evolution, but are not fully understood. 
 The number N$_{\rm C}$ of carbon atoms is strongly constrained by the ratios of the observed aromatic IR bands. 
The determination of the size distribution is made difficult by the complexity of the unknown PAH-emitting mixture (Appendix A.4), however.

The lower limit of the PAH size distribution is determined by UV photodissociation of the smallest PAHs. 
Photodissociation of interstellar PAHs has been modeled in many works since their discovery (Appendix A). It is agreed that carbon loss is much slower than aromatic
hydrogen loss. However, 
the estimate of the critical number of carbon atoms  below which total dehydrogenation occurs, N$_{\rm Ccr}$,  remains uncertain.  
We estimate in Appendix A.2 that a wide range must be considered for 
the value of N$_{\rm Ccr}$, from $\sim$35 to $\sim$50.
  
The uncertain modeling of the PAH size distribution (Appendix A.4)  agrees that the smallest PAHs, with N$_{\rm C}$ close to  N$_{\rm Ccr}$, represent a significant fraction of all PAHs (see Section 3). More precisely, 
if we define the total abundance of all PAHs with a given value of N$_{\rm C}$, $\chi$(N$_{\rm C}$)\,=\,n(N$_{\rm C}$)/n$_{\rm H}$, the order of 
magnitude of $\chi$(N$_{\rm C}$) should be $\sim$10$^{-8}$ for 
values of N$_{\rm C}$ close to N$_{\rm Ccr}$, that is,\ X$_{\rm C}$(N$_{\rm C}$) is about 1.5 10$^{-3}$ of the total interstellar carbon. In addition, 
the distribution of \ X$_{\rm C}$(N$_{\rm C}$)  should mildly decrease with N$_{\rm C}$.

This size distribution is the result of a rich photochemistry whose main processes are detailed in Appendices A.2 and A.3 and which are summarized in Fig.\ 1 
and Table A.1. PAH processing is dominated by frequent transient heating through UV photon absorption and by ionic accretion reactions, mainly H 
and O onto PAH cations and C$^+$ onto neutral PAHs. They may increase the 
value of N$_{\rm C}$ or decrease it by ejection of small hydrocarbon molecules C$_{\rm x}$H$_{\rm y}$ or CO. Sporadic UV heating favors the most stable isomers. The complexity of these macromolecules and of their chemical processes makes a precise modeling very difficult if not impossible.  Nevertheless, the relatively high abundances of PAHs with N$_{\rm C}$ close to N$_{\rm Ccr}$, which cannot be  fed from lower N$_{\rm C}$ values because growth of possible interstellar carbonaceous molecules with  N$_{\rm C}$\,$<$\,N$_{\rm Ccr}$ seems to be ineffective (Appendix A.4),  show the existence of a net downsizing rate from higher N$_{\rm C}$ values. 
This effective downsizing processing, which most likely results from sputtering from reactions with O and C$^+$, is needed to compensate for the eventual dehydrogenation of the smallest PAHs.
Its analysis (Appendix A.4) allows us to derive a rough estimate of the rate of formation of graphenic carbon clusters C$_{\rm n}$, which might subsequently efficiently transform into fullerenes, as we argue in the next 
sections.

\section{Formation of cages with N$_{\rm C}$\,$<$\,60  from dehydrogenated PAHs in the diffuse ISM}

This section describes the likely evolution of interstellar fully 
dehydrogenated graphene nanoflakes,  C$_{\rm n}$ 
with n\,$\sim$\,35-50, sporadically heated by UV-photon absorptions, into carbon cages, that is, \ close structures made of 2D carbon sheets, similar to 
fullerenes, but more irregular. We begin by first recalling the similar 
modeling of formation of interstellar C$_{60}$  from graphene sheets C$_{\rm n}$ with n slightly greater than 60, and second, the complex models 
of PAH H-photolysis and isomerization. We then argue that C$_{40-50}$ 
nanoflakes heated at a temperature close to 1000\,K may spontaneously form irregular carbon cages through the same succession of steps as are well documented for C$_{60}$, and eventually more stable fullerene isomers with the catalytic help of C$^+$ temporary accretion. 

\subsection{Top-down formation of interstellar C$_{60}$}

As we recalled, the formation of interstellar  C$_{60}$ is still debated (e.g., Candian et al.\ 2018). The most convincing propositions imply curling 
and closure of sheets of graphene that might be formed  from dehydrogenated interstellar PAHs in the ISM in the presence of strong UV radiation (Bern\'e et al.\ 2012, 2015a; Zhen et al.\  2014). These models are based on laboratory data and molecular dynamics simulations. The following two works are especially illuminating about key features of the possible formation of C$_{60}$ in the ISM: i) In situ high-resolution transmission electron microscopy performed by Chuvilin et al.\ (2010) revealed the formation of spheroidal carbon cages directly from graphene nanoflakes. ii) An accelerated molecular dynamics simulations in the range 300-3000\,K allowed Pietrucci \& Andreoni (2014) to perform a detailed investigation of the different steps of this process. 

These processes are inefficient in forming C$_{60}$ in the diffuse ISM because the  dehydrogenation of PAHs with N$_{\rm C}$\,$>$\,60 is impossible 
by single photon absorption. 
However, it is natural to wonder whether the same scheme might be applied 
to the formation of smaller fullerenes from fully dehydrogenated smaller PAHs, which are expected to be naturally produced in the diffuse ISM. This is the aim of this section.

\subsection{Cage formation from carbon nanoflakes}

Because of the striking success of the observation of the formation of C$_{60}$ from a graphene sheet in the laboratory by Chuvilin et al.\ (2010) 
and the detailed understanding of its various steps by the simulations of 
Pietrucci \& Andreoni (2014, hereafter PA14), it is worthwhile to analyze 
the possibility that the same processes may similarly proceed for forming smaller fullerenes with N$_{\rm C}$\,$\sim$\,40-50. We do this below for each 
of the steps from the detailed results of Pietrucci \& Andreoni and focus on their simulation  at 1000\,K because this temperature is relevant for PAHs with N$_{\rm C}$\,$\sim$\,N$_{\rm Ccr}$ in the diffuse ISM (Appendix A). 

$\bullet$ Initial graphene nanoflake.
As stated by PA14, representing the initial state of a fully dehydrogenated PAH by a pure plane graphene flake reproducing the PAH skeleton is certainly oversimplified. When heated to about 1000\,K by the absorption of 
a $\sim$10\,eV photon, it immediately reorganizes into a more stable form 
 with some general salient features, such as  
 edge  reorganization into a (5,7) rather than a (6,6) carbon pattern, or presence of at least one inner pentagon inducing a curved structure, such as corannulene. Although it is probable that  such structures do appear 
before the end of the total dehydrogenation (Parneix et al.\ 2017), this should not change the subsequent evolution of the flake much if it eventually becomes completely dehydrogenated. Nothing in these features should significantly change from N$_{\rm C}$\,=\,60 to N$_{\rm C}$\,=\,40.

$\bullet$ Formation of a convex or conical structure. The next step evolving 
toward a more complex structure to decrease the energy should also be general and little depend on whether there are 60 or 40 C. Its evolution  into a conical structure  with the formation of a 4C ring is  quite general  for N$_{\rm C}$\,=\,60  and the flake shape chosen by PA14.
We note, however, that the initial flake structure, with a bay, that was chosen by PA14 (their Fig.\ 1a), should favor the formation into a conical structure.
A similar evolution also appears likely for  N$_{\rm C}$\,$\sim$\,50, although this is less certain when N$_{\rm C}$ decreases strongly. In particular, the formation of a convex conical structure due to the presence of a four-membered ring can be hampered in smaller cages. The strain induced by the curvature might be too much there.

$\bullet$  Formation and fate of short carbon chains. 
Similarly,  the formation of various short chains, starting with C$_2$, appears quite general and independent of N$_{\rm C}$. However, a key question is their fate and the probability that they might detach, which is zero  for N$_{\rm C}$\,$\sim$60 in the PA14 simulations. It is striking in these simulations that every time a C$_2$ molecule detaches, it is reattached. The likelihood of reattachments for lower values of N$_{\rm C}$ is difficult to evaluate because their probability  depends on activation barriers and the efficiency of internal energy redistribution.  This might 
prevent the overall cage formation or at least shrink the size of fullerenes that are eventually formed. 
This point would need to be determined by systematical simulations similar to those of PA14. They would imply a considerable amount of computer time, however. 

$\bullet$ Cage closure. This final state transforms chain networks into regular cages, as observed by PA14   for C$_{60}$ simulations. It is expected to depend little on  N$_{\rm C}$. Once formed, the cage is expected to resist subsequent transient heatings by photon absorption well. It is expected to frequently exchange ionization states through the equilibrium between photoionization and electron recombination. Accretion of H, H$^+$, and e$^-$  is thought to be inefficient in the diffuse ISM because these particles should be immediately 
released by subsequent UV-photon absorptions.

Overall, the formation of cages from carbon nanoflakes with  N$_{\rm C}$\,$\sim$\,40-50 appears to be possible, especially for the highest values of  N$_{\rm C}$, but it is not guaranteed, even in a small proportion, because C$_2$ release might be dominant.  
In addition, it is not certain that the dehydrogenation of interstellar PAHs proceeds all the way to remove all H atoms. The possibility that the resulting carbon cage remains open and partially hydrogenated, eventually leading to tubular structures, might therefore deserve consideration.

\section{Cage evolution into fullerene}

\subsection{Fullerene cage evolution}

In their simulations of the evolution of a carbon nanoflake with 60 atoms 
at 1000\,K, PA14 found that the final state is always a cage with 60 carbon atoms. However, it is never the well-known most stable  $I_h$ C$_{60}$ 
isomer (buckminsterfullerene). Reaching the $I_h$ shape implies difficult 
cage rearrangements with high activation barriers, such as in classical Stone-Wales transformations of pentagon-heptagon structures into two hexagons (E$_{\rm A}$\,$>$\,6\,eV, see, e.g., Bettinger et al.\ 2003). Such isomerizations can be achieved at higher temperature, as observed at 3000\,K 
by PA14 (see also Bern\'e et al.\ 2015a). These temperatures are much higher than can be achieved by single-photon absorption in the diffuse ISM (Appendix A).

However, it is known that a high temperature may be avoided through bond-rearrangement catalysis produced by the adsorption of a carbon atom (Eggen et al.\ 1996, Ewels et al.\ 2002, Dunk et al.\ 2012). As shown by $^{13}$C studies (Christian et al.\ 1992; Dunk et al.\ 2012), the carbon adatom is scrambled with the C atoms of the cage through its incorporation into the C network, probably mostly by initial creation of a 7C ring. This lowers the activation energy for bond rearrangements by a large factor, $\sim$3-4, so that rearrangements favoring the most stable isomers can be achieved at T\,$\la$\,1000\,K. This appears to be a key explanation for the high efficiency of the formation of fullerenes, especially $I_h$-C$_{60}$ 
and $D_{5h}$-C$_{70}$, as stressed, for example, by Dunk et al.\ (2012). It is clear that this catalysis could operate for any fullerene and even for any $sp^2$ system, including PAHs.

We suggest that similar processes might favor the formation of the most stable fullerene isomers from the cages formed from dehydrogenated PAHs (Section 3.2). Rather than atomic C, the dominant process should be the accretion of C$^+$ on neutral cages. It should allow efficient bond rearrangement favoring the most stable cage isomers during subsequent transient 
heating by UV photon absorption.

The question is whether C$^+$ accretion might not  favor cage growth. The accretion of C$^+$ on neutral even-numbered cage might eventually lead to the formation of relatively stable close-shell odd-numbered cationic carbon cages.  However, odd-numbered cages must be much more unstable than 
even-numbered ones in the periods they become neutral through e-recombination. Therefore it is likely but not certain that the additional carbon is ejected following a subsequent UV-photon  absorption before the next C$^+$ accretion. We also note that the large predominance of even-numbered cages over odd-numbered ones is a constant feature of all fullerene formation experiments. 

In the diffuse ISM, C$^+$ accretion onto any neutral cage should proceed at a rate similar to PAHs of the same size, that is,\ $\sim$10$^{-4}$\,yr$^{-1}$ (Figure 1 and Table A.1). As discussed in the following section, this is probably significantly faster than any cage-sputtering process. We may therefore conclude that C$^+$ accretion might allow efficient bond rearrangement during subsequent transient heating by UV photon absorption. This could ensure that most cages formed with N$_{\rm C}$ atoms eventually terminate into the most stable fullerene isomer with the same N$_{\rm C}$.

\subsection{Fullerene sputtering and possible abundances}

Fullerenes are known to be very stable structures. This may be quantified 
by the enthalpy
change for the elimination of a C$_2$ molecule (the dominant fragmentation channel of C$_{\rm n}$ cages). For instance, Candian et al.\ (2019) give the values 8.51, 9.57, 11.18, and 10.24\,eV for the reaction C$_{\rm n}$ 
$\Rightarrow$ C$_{\rm n-2}$  $+$ C$_2$ with the most stable isomer of C$_{44}$, C$_{50}$, C$_{60}$, and C$_{70}$, respectively (see D{\'\i}az-Tendero et al.\ 2006 for a complete list). This is comparable to PAHs for C$_{44}$, but significantly higher for the heavier fullerenes, even C$_{50}$. 

For these values of the dissociation energy, $\ga$\,8\,eV, it is clear from the discussion of PAH dissociation of Appendix A  that the ejection of a C$_2$ molecule from a molecule with 40 to 50 C
cannot be achieved by the absorption of a single UV photon with E\,$<$\,13.6\,eV. It  requires the simultaneous absorption of two photons, which occurs only about every 10$^7$\,yr or more in the diffuse ISM (Fig.\ 1 and 
Table A.1). Other well-established processes for PAH sputtering by energetic  He$^+$ or e$^-$ are slightly slower (Micelotta 2010a,b and Fig.\ 1). 
It is possible  that the most efficient decay process for interstellar fullerenes in the diffuse ISM is the combined result of O and C$^+$ accretion, similarly to PAHs, but with important differences. The reaction of atomic oxygen with  fullerene cations could similarly yield the loss of one 
C atom through CO release, but this could depend on the fullerene stability because the result of this reaction is questionable even for C$_{60}$ (see references in Omont 2016). The result of C$^+$ accretion by neutral fullerenes could be more entangled: 
the integration of an additional  carbon in the network can favor isomerization, as quoted, and also compensate for carbon loss resulting from CO photolysis, but it could also favor C$_2$ photo-ejection.

A rough estimate of the eventual abundance of  each fullerene F$_{\rm i}$ from the balance between its formation and destruction can be formally derived. When we assume that fullerene  F$_{\rm i}$ forms from dehydrogenation of PAHs with about N$_{\rm Ccr}$ carbon atoms with a total probability $\eta_{\rm i}$,
the number density of F$_{\rm i}$, ${\rm n(F_i)}$, is derived from the equation 
\begin{equation}
{\rm  \eta_i \times   \gamma_{spPAH} \times n_{PAHcr} =  \gamma_{spFi} \times n(F_i)}
,\end{equation} 

where ${\rm n_{PAHcr}}$ is the  number density of PAHs just above  N$_{\rm Ccr}$, $\gamma_{\rm spPAH}$ is their downsizing rate, which brings them below   N$_{\rm Ccr}$ (Appendix A.4) and  $\gamma_{\rm spFi}$ is the decay rate of ${\rm n(F_i)}$ (Appendix B.3). Therefore, we may write 
\begin{equation}
{\rm n(F_i) = \eta_i \times r_{spi} \times n_{PAHcr} }
,\end{equation}
where r$_{\rm spi}$\,=\,${\rm \gamma_{spPAH}/\gamma_{spFi}}$ is the ratio of the decay rates. In Appendix A.4 we estimate the value of the fraction of interstellar carbon in PAHs for each value of N$_{\rm C}$ above and close to N$_{\rm Ccr}$, ${\rm X_{C}}$\,$\sim$\,1.5$\times$10$^{-3}$ (i.e.,\ about 1\,\% of the total carbon in all PAHs). This yields for the carbon fraction in fullerene i
\begin{equation}
{\rm X_{CFi} = 1.5 \times 10^{-3} \times \eta_i \times r_{spi}}
.\end{equation}

 In the case of the most stable fullerenes, we may expect  r$_{\rm spi}$\,$\ge$\,1   because their stability is greater than that of PAHs, while  r$_{\rm spi}$ might be closer to 1 
for unstable fullerene isomers (see, e.g., Candian et al.\ 2019; Sun et al.\ 2005 and Table B.2 for an estimate of the formation energy of various 
fullerene isomers). If $\eta$ is the total probability that PAH dehydrogenation ends in  fullerenes, 
they should contain at least a total fraction ${\rm X_{CF}}$\,$\sim$\,$\eta  ~ \times$ 1.5 $\times$ 10$^{-3}$ of interstellar carbon. 
The value of $\eta$ remains very uncertain. It might be very small if dehydrogenated PAHs were mainly photodissociated through successive C$_2$ ejections (or formed nanotubes). Nevertheless, if 
they form efficiently from dehydrogenated PAHs, some of these fullerenes 
might have an abundance comparable to or greater than C$_{60}^+$, whose abundance in the diffuse ISM may be roughly estimated as $\sim$2 $\times 10^{-4}$ from its DIB strength (Walker et al.\ 2015; Campbell \& Maier 2017; Omont 2016). Large abundances like this, if confirmed,  should make some fullerenes smaller than C$_{60}$ detectable through their IR emission 
bands or their visible absorption bands (DIBs), as we discuss in the next 
section.

\section{Optical and infrared spectra}

\subsection{Mid-infrared spectrum and abundance limit}

In addition to the identification of  C$_{60}^+$ DIBs,  all detections of C$_{60}$ as well as  C$_{60}^+$  and  C$_{70}$ in the ISM have  been achieved 
through their mid-IR emission bands in {\it Spitzer} 6-20\,$\mu$m spectra 
(Cami et al.\ 2010; Sellgren et al.\ 2010; Bern\'e et al.\ 2015a,b, 2017; 
and, e.g., references in Cami 2014; Candian et al.\ 2018). The possible presence of smaller (44-, 50-, and 56-atom) fullerenes 
was recently addressed by Candian et al.\ (2019). Their conclusion is that the spectra of most cages that
are smaller than C$_{60}$  show features in the 13-15\,$\mu$m range, where the astronomical spectra of fullerene-rich planetary nebulae also contain characteristic signals. They were unable to conclusively identify any of them in {\it Spitzer} spectra, however, even the most symmetrical $C_{5h}$ C$_{50}$, because the IR spectra of their various isomers are similar and the strength of their IR features is weaker than C$_{60}$.

The same conclusions apply for the {\it Spitzer} IR spectra of diffuse regions of the ISM displayed by Bern\'e  et al.\ (2017). Comparing these spectra with those computed for the different isomers of C$_{50}$ and C$_{44}$ shown by Candian et al.\ (2019), we may conclude that none of them and their cations may have an abundance significantly greater than that of C$_{60}$, which is estimated as 2 $\times 10^{-4}$ at most of the total 
interstellar carbon by Bern\'e  et al.\ (2017). However, abundances close 
to this limit are not completely excluded. They should be more easily detectable by the James Webb Space Telescope (JWST). 

\subsection{Fullerene optical spectra }

\begin{figure}[htbp]
         \begin{center}
\includegraphics[scale=0.55, angle=0]{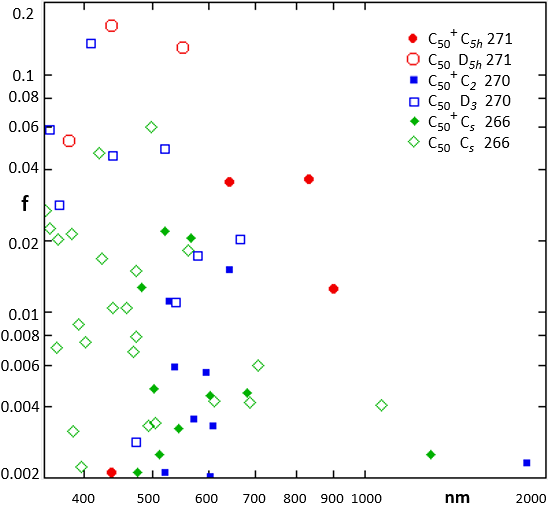}
 \caption{Computed absorption spectra (f-values) of three isomers of  fullerene C$_{50}$ (Table B.2).
The results of complete calculations from $\sim$350\,nm to 2000\,nm for the isomers of neutral C$_{50}$ ($D_{5h}$, 271; $D_3$, 270; and $C_s$, 266) are displayed by open symbols, together with the cation spectra ($C_{5h}$, 271; $C_2$, 270; and $C_s$, 266) from $\sim$500\,nm to 2000\,nm (full symbols).
See similar Figures B.1 to B.4 for isomers of C$_{40}$, C$_{42}$, C$_{44}$, C$_{46}$, and C$_{48}$ listed in Table B.2.}
     \end{center}
 \end{figure}

The visible range provides an incomparable sensitivity to detect absorption lines on the sightline of bright stars. About 600 such DIBs are known, 
whose mostly unknown carriers  are believed to be carbonaceous macromolecules, such as the four confirmed C$_{60}^+$ DIBs. However, even the identification of the carriers of the strongest DIBs needs high-accuracy laboratory spectral data, as is shown by the history of the confirmation of the identification of the C$_{60}^+$ DIBs (Foing \& Ehrenfreund 1994; Campbell 
et al.\ 2015). 

Fullerenes and their cations are known to have only relatively weak bands in the visible range (Koponen et al.\ 2008; Lan, Kang \& Niu 2015). 
However, no  detailed laboratory or theoretical data about the optical spectrum of fullerenes except for C$_{60}$ and C$_{70}$ are available, including those with N$_{\rm C}$ in the range 40-50, which we propose are the most likely possible products of PAH dehydrogenation.  Therefore we performed density functional theory (DFT) B3LYP/6-31G* calculations 
of the visible spectrum for 19 neutral fullerene isomers (and their cations, except for one), with even values of N$_{\rm C}$ between 40 and 50 (Table B.2).  The computation methods  are described and analyzed in detail in Appendix B.1. 

The results are presented in Figs.\ 2 and B.1 to B.4.
All cases display at least one and generally several significant bands between 400 and 800 nm, with f-values higher than 10$^{-2}$. There is a trend for finding stronger bands for the most stable fullerenes, C$_{50}$ and C$_{44}$, and increasing values of N$_{\rm C}$. C$_{50}$ isomers have several visible bands with f $\sim$ (5-8)$\times$10$^{-2}$. Most of these relatively strong bands are concentrated at $\lambda$ $\la$ 550 nm, but a 
few are found at $\sim$550-600 nm (Fig.\ 2). The visible bands of the isomers of the other fullerenes display a similar wavelength distribution, but with significantly lower f-values, by a factor $\sim$2 for C$_{44}$ and C$_{48}$, $\sim$3 for C$_{46}$, and  $\sim$6 for C$_{40}$ and  C$_{42}$ (see Figures B.1 to B.4). 
Our calculations  generally confirm some similarity of the spectra of cation and neutral forms of the same molecule, as is known for PAHs (e.g., Malloci et al.\  2004).

The extension of the calculations to C$_{60}^+$ with the same methods has
produced a strong double band at 893/892 nm with both f-values of 0.032, to be compared with the actual values of 963.2 and 957.7 nm with f\,$\sim$\,0.03-0.02 (Walker et al.\ 2015; Campbell \& Maier 2017). We may therefore conclude that i) the order of magnitude of the calculated f-values appears to be good, and ii) the most stable fullerene isomers with N$_{\rm C}$\,$\sim$40-50 generally display one strong band at least in the visible-near-IR range comparable to C$_{60}^+$.

\subsection{Fullerenes as possible DIB carriers}

Eq.\ (1) shows  that fullerenes with an f-value of a few 10$^{-2}$ and an abundance, X$_{\rm Ci}$, up to a few 10$^{-4}$,  may carry medium-strength DIBs with an equivalent width EW$_{\rm i}$ in the range 10-100\,m\AA /mag. This could make fullerenes with N$_{\rm C}$\,$\sim$\,40-50 significant DIB 
carriers if they were able to form from dehydrogenated PAHs, with individual abundances reaching X$_{\rm Ci}$\,$\sim$\,(1-2)$\times$10$^{-4}$. We note, however, that the mid-IR spectrum of diffuse interstellar clouds appears to preclude larger fullerene abundances (Section 5.1) and that the number of species with X$_{\rm Ci}$\,$\sim$\,10$^{-4}$ is probably limited in any case.

Figs.\ 2 and B.1 to B.4 show that the DIBs carried by such neutral and cation fullerenes might be distributed across the whole DIB wavelength range from 400 to 900\,nm, with a significant number of cation bands at $\lambda$\,$>$\,600\,nm.
As quoted, such abundances of 10$^{-4}$ or slightly higher are comparable to those estimated in diffuse clouds for C$_{60}^+$ and C$_{60}$ from near-IR DIBs and mid-IR emission bands, and larger than that of C$_{70}^+$. Despite the poorer stability of the fullerenes with N$_{\rm C}$\,$\sim$\,40-50, it is not impossible that their abundance is comparable to C$_{60}$ 
because C$_{60}$ 
cannot be directly formed from photodissociated PAHs in diffuse clouds. Therefore it is possible that these fullerenes carry some DIBs, especially cations such as C$_{50}^+$ and C$_{44}^+$ at longer wavelength. Their intensity could be somewhat similar to the weakest C$_{60}^+$ DIBs (Walker 
et al.\ 2016). Nothing is known about the width of their visible bands, but it might be comparable with the narrow width (a few cm$^{-1}$) measured in the gas-phase for C$_{60}^+$ (Campbell et al.\ 2015) and C$_{60}$ (Close et al.\ 1997) and in the ISM for C$_{60}^+$ DIBs (Cox et al.\ 2014), which could make them relatively easy to detect as DIBs. 

The mid-IR $Spitzer$ data available on the diffuse ISM (Bern\'e et al.\ 2017) place a limit on the total abundance of such fullerenes with N$_{\rm C}$\,$\sim$\,40-50, which we estimate to be on the order of X$_{\rm Ctot}\,\la\,10^{-3}$ (Section 5.1). When we assume an average oscillator strength  ${\rm \bar{f}}$\,$\sim$\,0.03, we obtain a total equivalent width EW$_{\rm ful}$\,$\la$\,300 m\AA /mag from Eq.\ (1) for all these fullerenes. This is only about 3\% of the total equivalent width of the sum of all known DIBs (but $\lambda$4429\AA\ and $\lambda$7700\AA \footnote{Extremely strong and broad DIB recently discovered by  Ma{\'\i}z Apell{\'a}niz et al.\ (2021).}) in the diffuse ISM, which is estimated to be $\sim$11000 m\AA /mag in a typical sightline (Hobbs et al.\ 2009, Omont \& Bettinger 2020). This confirms the conclusion of Omont (2016) that fullerenes can be only a minor component of the DIB carriers. 

These conclusions again show that the condition derived from Eq.\ 
(1) for the total abundance of all DIB carriers is extremely constraining. Assuming ${\rm \bar{f}}$\,$\sim$\,0.1, as proposed by Salama et al.\ (1996) for PAH-cation DIB carriers, yields X$_{\rm Ctot}\,\sim\,10^{-2}$ for the carriers of all DIBs (but $\lambda$4429\AA\ and $\lambda$7700\AA). This is a significant fraction, $\sim$10\%, of the total carbon included in PAHs (but less than that if f\,$>$\,0.1). It is very difficult indeed to escape the conclusion that the bulk of the DIB carriers must be related to the PAHs. Except for fullerenes, the only contenders, carbon chains 
(or rings), have never been identified through their mid-IR emission, which makes this high abundance unlikely. In addition to their overwhelming abundance, PAHs have many advantages as DIB carriers, including relatively strong bands at short wavelength (e.g.,\ Malloci et al.\ 2007) and even at long $\lambda$ for cations; a possible enhanced abundance for symmetric pericondensed PAHs  (Andrews et al.\ 1995); families with very strong bands (f\,$\ga$\,1), such as polyacenes and rylenes (Omont et al.\ 2019; Ruiterkamp et al.\ 2002), which are favorable for carrying C$_2$ DIBs and other narrow DIBs (Omont \& Bettinger 2020). Their extreme isomer variety and possible large band width can prevent their individual detection, however, especially at 
short $\lambda$, which might explain the low DIB density at $\lambda\,<$\,5500\,\AA .

Other DIB carriers remain possible, including nanotube segments and hetero atoms inserted in these various carbonaceous macromolecules. Nanotube segments present similarities with symmetrical PAHs and fullerenes. Their strong visible bands justify their consideration as DIB carriers (Zhou et al.\ 2006).  As their formation appears to be possible from partially 
dehydrogenated PAHs (Section 3.2; Chen et al.\ 2020), we calculated the optical absorption spectra of a few of the shortest neutral and cation (4,4) and (5,5) hydrogenated armchair tube-segments with the same method as for fullerenes (Appendix B.2). Figs.\ B.6 and B.7 show that all such neutral and cation tube-segments with N$_{\rm C}\,>$\,40 have a strong band in the range $\sim$400-800\,nm with f\,$\sim$\,0.5. There are many 
additional weaker bands, including cation bands at $\lambda$\,$\ga$\,550\,nm with f\,$\ga$\,0.02.  As discussed in Appendix B.2, the size distribution of  such nanotube-segments and the resulting interstellar absorption spectrum might depend not only on their formation, but also on their possible subsequent growth by C$^+$accretion. 
 As it is  expected that their band width is similar to that of PAHs or fullerenes, it is confirmed that they should also be considered as DIB carriers, as suggested by Zhou et al.\ (2006), if they actually formed in the diffuse ISM.

As the formation of semicapped nanotube segments from partially dehydrogenated PAHs  is not excluded,  we extended the calculations to a few  semicapped  armchair (5,5) nanotube segments, C$_{\rm n}$H$_{10}$, with n\,=\,50 to 100, including C$_{50}$H$_{10}$, which was synthesized by Scott 
et al.\ (2011). 
Their visible absorption spectra (Fig.\ B.8) have similar general characteristics to open nanotube segments, with a trend to slightly weaker features.

\section{Conclusion}
The high interstellar abundance of PAHs and their size distribution are the result of complex chemical processes implying dust, UV radiation, and the main  gaseous components (H, C$^+$, O, etc.). 
A downsizing process, possibly dominated by sputtering following atomic oxygen accretion onto PAH cations,
is required to compensate for  efficient C$^+$ accretion and to maintain a 
high C$^+$ gaseous abundance. 
These processes must explain the high abundance of relatively small PAHs in the diffuse interstellar medium. They imply the continuous  formation of some PAHs that are small enough (number of carbon atoms N$_{\rm C}$ $\la$ 35-50) to be completely dehydrogenated by interstellar UV radiation. 

The carbon clusters C$_{\rm NC}$ thus formed are constantly exposed to the absorption of $\sim$10-13.6\,eV UV photons, heating them to a temperature 
close to 1000\,K, which allows isomerization and favors the formation of the most stable isomers with curved shapes.  They might tend to form irregular carbon cages.
The frequent accretion of interstellar C$^+$ ions followed by UV photon absorption might favor further cage isomerization as known in the laboratory for C$_{60}$, possibly yielding the formation of the most stable fullerenes, such as C$_{40}$, C$_{44}$, and C$_{50}$. 

These fullerenes are expected to be very stable in the diffuse ISM because C$_2$ ejection is not possible by single UV photon absorption but might need rare two-photon absorption, while the increase in the size  of the fullerene cages is impeded by the rarity of small carbonaceous molecules C$_{\rm x}$H$_{\rm y}$ with x\,$\ge$\,2. 
It is therefore possible that at least one of these fullerenes or its cation is as abundant as C$_{60}$ or C$_{60}^+$ in the diffuse ISM, although 
higher abundances are precluded by the lack of observed  matching mid-IR features. 
Our B3LYP calculations of the visible spectrum for a score of fullerene isomers  with  N$_{\rm C}$ between 40 and 50 confirmed that they generally have a few spectral bands in the visible range with f-values in the range of a few 10$^{-2}$. 
This could make some of them interesting candidates for the carriers of  diffuse interstellar bands. 

As is well known, the eventual identification of additional fullerenes as carriers of relatively weak DIBs cannot be achieved without laboratory measurement of the gas-phase wavelengths of their visible spectral bands. Even with the progress of methods such as He-tagging for cations, these measurements imply a considerable experimental effort, including the synthesis of such fullerenes. Nevertheless, this effort may be justified because it is important to understand the unique chemical properties of interstellar carbonaceous nanoparticles and macromolecules. Dehydrogenated PAHs are certainly the main source of interstellar carbon clusters with a few dozen carbon atoms. The evolution of these clusters is a key question in interstellar chemistry.
Fullerenes appear to be a possible logical step in this evolution. If confirmed, their possible identification as DIB carriers would provide a powerful and specific tool to further reveal their detailed interstellar abundances and the astrochemical processes that determine them. 

Finally, if the dehydrogenation of the smallest interstellar PAHs is not complete, it might lead to the formation of hydrogenated nanotube segments that are either open at both ends or are semicapped. 
If this formation appears to be more difficult than that of fullerenes, this possibility should also be explored.
As  most of these tubular PAHs with a few dozen carbon atoms have at least a strong visible band, with an expected width similar to PAHs or fullerenes, they should also be considered as DIB carriers.

\bigskip
\begin{acknowledgements}
We thank the referee for his/her very helpful comments and suggestions. 
We are indebted 
to Pierre Cox for his careful reading of the manuscript and his suggestions for improving it. 
We made a high use of "TOPCAT \& STIL: Starlink Table/VOTable Processing Software" (Taylor 2005) and thank its developers.
The authors acknowledge support by the state of Baden-Württemberg through bwHPC
and the German Research Foundation (DFG) through grant no INST 40/575-1 FUGG (JUSTUS 2 cluster). 
\end{acknowledgements}

\bigskip

{\bf References}  
\bigskip

Adamo, C.\ \& Barone, V.~J.\ 1999, Chem.\ Phys., 110, 6158

Allamandola, L.~J., Tielens, A.~G.~G.~M., \& Barker, J.~R.\ 1985, \apjl, 290, L25

Allamandola, L.~J., Tielens, A.~G.~G.~M., \& Barker, J.~R.\ 1989, \apjs, 71, 733

Andersson, K.\ 2020, Chem.\ Phys.\ Lett., 739, 136976

Andrews, H., Boersma, C., Werner, M.~W.\ et al.\ 2015, \apj, 807, 99

Andrews, H., Candian, A., \& Tielens, A.~G.~G.~M.\ 2016, \aap, 595, A23

Becke, A.~D.~J.\ 1993, Chem.\ Phys., 98, 5648

Bernal, J.~J., Haenecour, P., Howe, J., et al.\ 2019, \apjl, 883, L43

Bern{\'e}, O. \& Tielens, A.~G.~G.~M.\ 2012, Proceedings of the National Academy of Science, 109, 401

Bern{\'e}, O., Montillaud, J., \& Joblin, C.\ 2015a, \aap, 577, A133

Bern{\'e}, O., Montillaud, J., Mulas, G., et al.\ 2015b, SF2A-2015: Proceedings of the Annual meeting of the French Society of Astronomy and Astrophysics, 65

Bern{\'e}, O., Cox, N.~L.~J., Mulas, G., et al.\ 2017, \aap, 605, L1

Bettinger, H.~F., Yakobson, B.~I.\ \& Scuseria, G.~E.\ 2003, J.\ Am.\ Chem.\ Soc., 125, 5572

Cami, J., Bernard-Salas, J., Peeters, E., \& Malek, S.~E.\ 2010, Science, 
329, 1180 

Cami, J.\ 2014, IAU Symposium, 297, 370 

Cami, J., \& Cox, N.~L.~J.\ 2014, IAU Symposium, 297

Cami, J., Peeters, E., Bernard-Salas, J., et al.\ 2018, Galaxies, 6, 101

Campbell, E.~K., Holz, M., Gerlich, D., \& Maier, J.~P.\ 2015, \nat, 523, 
322 

Campbell, E.~K., \& Maier, J.~P., 2017, J.\ Chem.\ Phys., 146, 160901

Candian, A., Zhen, J., \& Tielens, A.~G.~G.~M.\ 2018, Physics Today, 71, 38

Candian, A., Gomes Rachid, M., MacIsaac, H., et al.\ 2019, \mnras, 485, 1137

Canosa, A., Laub\'e, S., Rebrion, C., Pasquerault, et al.\ 
1995, Chem.\ Phys.\ Lett., 245, 407

Castellanos, P., Candian, A., Andrews, H., \& Tielens, A.~G.~G.~M.\ 2018a, \aap, 616, A166 

Castellanos, P., Candian, A., Zhen, J., Linnartz, H., \& Tielens, A.~G.~G.~M.\ 2018b, \aap, 616, A167

Chai, J.-D.\ \& Head-Gordon, M.~J.\ 2008a, Chem.\ Phys., 128, 084106

Chai, J.-D.\ \& Head-Gordon, M.~J.\ 2008b, Phys.\ Chem.\ Chem.\ Phys., 10, 6615

Chen, T., Zhen, J., Wang, Y., Linnartz, H., \& Tielens, A. G. G. M. 2018, 
CPL,
692, 298

Chen, T., Luo, Y., \& Li, A.\ 2020, \aap, 633, A103. doi:10.1051/0004-6361/201936873

Christian, J.F.\ et al.\ 1992c, J.\ Phys.\ Chem., 96, 3574

Chuvilin, A.\ et al.\ 2010, Nature Chemistry, 2, 450

Close, J.D., Federmann, F., Hoffmann, K., Quaas, N.\ 1997, Chem.\ Phys.\ Lett., 276, 393

Cox, N.~L.~J.\ 2011, The Molecular Universe, 280, 162. doi:10.1017/S1743921311024951

Cox, N.~L.~J., Cami, J., Kaper, L., et al.\ 2014, \aap, 569, A117

Demarais, N.~J., Yang, Z., Snow, T.~P., et al.\ 2014, \apj, 784, 25. doi:10.1088/0004-637X/784/1/25

Desert, F.-X., Boulanger, F., \& Puget, J.~L.\ 1990, \aap, 500, 313

D{\'\i}az-Tendero, S., S{\'a}nchez, G., Alcam{\'\i}, M., et al.\ 2006, International Journal of Mass Spectrometry, 252, 133

Draine, B.~T. \& Li, A.\ 2007, \apj, 657, 810

Draine, B.~T., Li, A., Hensley, B.~S., et al.\ 2020, arXiv:2011.07046

Dunk, P.~W., Kroto, H.~W.\ et al.\  2012, Nat. Commun., 3, 855.

Eggen, B.~R., Heggie, M.~I., Jungnickel, G., et al.\ 1996, Science, 272, 87

Ernzerhof, M.\ \& Scuseria, G.~E.~J.\ 1999, Chem.\ Phys., 110, 5029

Ewels, C.~P., Heggie, M.~I.\ \& Briddon, P.~R.\ 2002, Chem.\ Phys.\ Lett., 351, 178

Foing, B.~H., \& Ehrenfreund, P.\ 1994, \nat, 369, 296

Fowler, P.~W.\ \& Manolopoulos, D.~E.\ 1995, An Atlas of Fullerenes, Oxford University Press, Oxford. 

Frisch, M.~J., Trucks, G.~W., Schlegel, H.~B., Scuseria, G.~E.\ et al.\ 2016, Gaussian 16 Rev.\ C.01, Wallingford, CT

Fulara, J., Lessen, D.,  Freivogel, P., \& Maier, J.~P.\ 1993a, \nat, 366, 439 

Gredel, R., Carpentier, Y., Rouill{\'e}, G., et al.\ 2011, \aap, 530, A26

Hobbs, L.~M., York, D.~G., Thorburn, J.~A., et al.\ 2009, \apj, 705, 3

Joblin, C. \& Tielens, A.~G.~G.~M.\ 2011, EAS Publications Series, 46

Koponen, L., Puska, M.~J.\ \& Nieminen, R.\~M.\ 2008, J.\ Chem.\ Phys., 128, 154307

Krasnokutski, S.~A., Huisken, F., J{\"a}ger, C., et al.\ 2017, \apj, 836, 
32. doi:10.3847/1538-4357/836/1/32

Lan, Y., Kang, H.\ \& Niu, T.\ 2015, Eur.\ Phys.\ J.\ D, 69, 69

Leach, S.\ 1996, Z. Physikalische Chemie, 195, 15

Lee, C., Yang, W.\ \& Parr, R.~G.\ 1988, Phys.\ Rev., B,  37, 785

Leger, A. \& Puget, J.~L.\ 1984, \aap, 500, 279

Le Page, V., Snow, T.~P., \& Bierbaum, V.~M.\ 2001, \apjs, 132, 233 

Le Page, V., Snow, T.~P., \& Bierbaum, V.~M.\ 2003, \apj, 584, 316 

Lepp, S. \& Dalgarno, A.\ 1988, \apj, 324, 553

Li, A. \& Draine, B.~T.\ 2001, \apj, 554, 778

Li, J., Kudin, K.~N.,  McAllister, M.~J., Prud’homme, R.~K., Aksay, I.~A.\ \& Car, R.\ 2006, Phys.\ Rev.\ Lett., 96, 176101

Ma{\'\i}z Apell{\'a}niz, J., Barb{\'a}, R.~H., Caballero, J.~A., et al.\ 2021, \mnras, 501, 2487. doi:10.1093/mnras/staa2371

Malloci, G., Mulas, G., \& Joblin, C.\ 2004, \aap, 426, 105. doi:10.1051/0004-6361:20040541

Malloci, G., Joblin, C., \& Mulas, G.\ 2007, Chemical Physics, 332, 353. doi:10.1016/j.chemphys.2007.01.001

Maragkoudakis, A., Peeters, E., \& Ricca, A.\ 2020, \mnras, 494, 642. doi:10.1093/mnras/staa681

McGuire, B.~A., Loomis, R.~A., Burkhardt, A.~M., et al.\ 2021, Science, 371, 1265. doi:10.1126/science.abb7535

Micelotta, E.~R., Jones, A.~P., \& Tielens, A.~G.~G.~M.\ 2010a, \aap, 510, A36 

Micelotta, E.~R., Jones, A.~P., \& Tielens, A.~G.~G.~M.\ 2010b, \aap, 510, A37 

Montillaud, J., Joblin, C., \& Toublanc, D.\ 2013, \aap, 552, A15 

Omont, A.\ 1986, \aap, 164, 159 

Omont, A.\ 2016, \aap, 590, A52 

Omont, A., Bettinger, H.~F., \&  T\"{o}nshoff, C.\ 2019,  \aap, 625, 441 

Omont, A. \& Bettinger, H.~F.\ 2020, \aap, 637, A74. doi:10.1051/0004-6361/201937071

Parneix, P., Gamboa, A., Falvo, C., et al.\ 2017, Molecular Astrophysics, 
7, 9

Pietrucci, F., \& Andreoni, W.\ 2014, Journal of Chemical Theory and Computation, 10, 913  (PA14)

Pilleri, P., Herberth, D., Giesen, T.~F., et al.\ 2009, \mnras, 397, 1053. doi:10.1111/j.1365-2966.2009.15067.x

Series, 46, 355. doi:10.1051/eas/1146037

Puget, J.~L. \& Leger, A.\ 1989, \araa, 27, 161

Ruiterkamp, R., Halasinski, T., Salama, F., et al.\ 2002, \aap, 390, 1153

Salama, F., Bakes, E.~L.~O., Allamandola, L.~J., \& Tielens, A.~G.~G.~M.\ 
1996, \apj, 458, 621 

Scott, L.~T.,  Jackson, E.~A., Zhang, Q., Steinberg, B.~D., Bancu, M.\ \& 
Li, B.\ 2011, J.\ Am.\ Chem.\ Soc., 134, 107

Sellgren, K., Werner,  M.~W., Ingalls, J.~G., et al.\ 2010, \apjl, 722, L54 

Snow, T.~P., \& Bierbaum, V.~M. 2008, Annu.\ Rev.\ Anal.\ Chem., 229

Steglich, M., Bouwman, J., Huisken, F., \& Henning, T.\ 2011, \apj, 742, 2 

Sun, G., Nicklaus, M.~C.\ \& Xie, R.\ 2005, J.\ Phys.\ Chem.\ A, 109, 4617

Taylor, M.~B.,  2005, Astronomical Data Analysis Software and Systems XIV, eds. P.\ Shopbell et al., ASP Conf. Ser. 347, 29

Thaddeus, P.\ 1995, in Tielens, A.~G.~G.~M., \& Snow, T.~P.\ 1995, The Diffuse Interstellar Bands, {\it Astrophysics and Space Science Library}, 202, Kluwer

Tielens, A.~G.~G.~M.\ 2005, The Physics and Chemistry of the Interstellar 
Medium (Cambridge University Press, Cambridge, UK)

Tielens, A.~G.~G.~M.\ 2008, \araa, 46, 289 

Tielens, A.~G.~G.~M.\ 2013, Reviews of Modern Physics, 85, 1021


Walker, G., Bohlender, D., Maier, J., \& Campbell, E.\ 2015, \apjl, 812, L8  

 Walker, G.~A.~H., Campbell, E.~K., Maier, J.~P., Bohlender, D., \& Malo, 
L.\ 2016, \apj, 831, 130 

Yanai, T., Tew, D.~P.\ \& Handy, N.~C.\ 2004, Chem.\ Phys.\ Lett., 393, 51

Zhao, Y.\ \& Truhlar, D.\ 2008, Theor.\ Chem.\ Acc., 120, 215

Zhen, J., Castellanos, P., Paardekooper, D.~M., et al.\ 2014, \apjl, 797, 
L30. doi:10.1088/2041-8205/797/2/L30

Zhen, J., Rodriguez Castillo, S., Joblin, C., et al.\ 2016, \apj, 822, 113 

Zhou, Z., Steigerwald, M., Hybertsen, M., Brus, L.\ \& Friesner, R.~A.\ 2004, J.\ Am.\ Chem.\ Soc.\ 126, 3597

Zhou, Z., Sfeir, M.~Y., Zhang, L., Hybertsen, M.~S., Steigerwald, M.\ \&  
Brus, L.\ 2006, \apjl, 638, L105 

\appendix  {}

\section{Summary of the properties and chemistry of the smallest interstellar 
PAHs in the diffuse ISM}

\subsection{A main component of the ISM}

The PAHs are a key component of the ISM because they are the main reservoir of carbonaceous macromolecules and typically contain $\sim$10\% of carbon in the diffuse ISM. Since their identification  (L\'eger \& Puget 1985; Allamandola et al.\ 1985), they have been the object of many studies that  interpreted  their IR emission and discussed their physical and chemical properties (see, e.g., reviews by Omont 1986; Puget \& L\'eger 1989; Allamandola et al.\ 1989;  Salama et al.\ 1996; Li \& Draine 2001; Tielens 2008, 2013; Joblin \& Tielens 2011; Candian et al.\ 2018).

Practically our entire knowledge about astrophysical PAHs  is based on their series of emission features in between 3.3-20\,$\mu$m, which may be modeled as complex fluorescence following the absorption of a UV photon. The observed spectrum always results from a complex mixture of many PAHs with N$_{\rm C}$ $\sim$ 50-150 and provides key information about their size distribution (Sect.\ 2.3),  their charge state, and also, with more difficulty, about their possible dehydrogenation and superhydrogenation. 

Because of their large photoionization cross-section,  interstellar  PAHs always contain a substantial fraction of cations. This fraction depends both on the UV intensity and the gas density. In typical conditions of the diffuse 
ISM, PAHs undergo very frequent changes between neutral and cation charge 
states at a rate close to 1 yr$^{-1}$ (see Fig.\ 1 and Table A.1). Any discussion of PAH chemistry must therefore jointly address neutrals and cations.

The overall formation or destruction processes of interstellar PAHs are complex. Their initial formation  takes place mainly in circumstellar shells of AGB carbon stars, which eject a large amount of carbon dust into the ISM. 
Part of this dust may be converted into PAHs in shocks in the post-AGB or 
planetary nebula stages. A similar formation of PAHs may take place in supernova remnants. However, most of the observed PAHs are thought to result from the complex processing of dust grains and PAHs in the ISM through C$^+$/C accretion, various sputtering processes, accretion onto dust, and shock processing. 
PAH destruction  ultimately occurs through  energetic processes including 
mostly He$^+$ and e$^-$  (Micelotta et al.\ 2010a,b), and possibly through O sputtering (see below). In the next section, we review the properties and chemistry of the smallest interstellar PAHs  in the diffuse ISM. This is relevant for their eventual degradation into fullerenes.

\subsection{Photo-processes}

The chemistry of interstellar PAHs has been discussed by many authors (e.g., Omont (1986); Lepp \& Dalgarno (1988); Le Page et al.\ (2001, 2003); Tielens (2008, 2013)). It is dominated by photochemistry, as we discuss now, and ionic accretion processes (Section A.3), with rates whose orders of magnitude are recalled in Table A.1.

$\bullet$ UV absorption. This is the most frequent and the key process for the chemical evolution of interstellar PAHs, even in the mild conditions of 
the diffuse ISM. At a frequency of about one month, at a rate that is practically proportional to N$_{\rm C}$ and independent of the charge and hetero atoms, it injects an energy of $\sim$6-13.6\,eV, which may produce photoionization, photodissociation, or structure reorganization. As most of the photon energy is radiated in less than 1\,s, the rate of simultaneous absorption of two 
UV photons, which may degrade the PAH carbon skeleton, is typically $\la$10$^{-7}$\,yr$^{-1}$.

$\bullet$ Photoionization or electron recombination. If the energy of the photon is significantly greater than the PAH first-ionization potential, I$_{\rm P}$\,$\sim$\,6.5\,eV, the probability of photoionization before internal thermalization is significant. As electronic recombination occurs at a similar but uncertain rate $\sim$1\,yr$^{-1}$, the fractions of neutrals and cations are both significant.

$\bullet$ Photodissociation. 
H photolysis is the main reaction outcome from UV  absorption. The photon 
energy E$_{\rm UV}$
is mostly rapidly thermalized, yielding a vibrational temperature, T$_{\rm V}$(K)\,$\sim$\,2000$\times$(E$_{\rm UV}$(eV)/N$_{\rm C})^{0.4}$ (Tielens 2005), close to 1000\,K for E$_{\rm UV}$\,$\sim$\,10\,eV and N$_{\rm C}\,\sim$\,50. Before this energy is progressively reemitted in the infrared within less than a fraction of a second, every possible reorganization can take place, either isomerization with an activation energy E$_{\rm A}$ lower than a few eV, or photolysis with E$_{\rm A}$\,$\la$\,5\,eV.  Both processes critically depend on T$_{\rm V}$ and therefore on N$_{\rm C}$. 

Because of the exponential dependence on T$_{\rm V}$, the dehydrogenation 
of PAHs is expected to vary extremely sharply  around a critical value of N$_{\rm 
C}$,  N$_{\rm Ccr}$, so that the degree of hydrogenation is expected to abruptly vary from complete to zero around N$_{\rm Ccr}$. 
We note, however, that these PAHs may  retain their last few hydrogens, as quoted by Castellanos et al.\ (2018a). 
With the typical UV spectrum of the diffuse ISM, the value of N$_{\rm Ccr}$ probably depends little on the values assumed in Table A.1 for n$_{\rm H}$, T, and G$_0$. 
Its value remains uncertain, however, because of the variety of PAH isomers and 
the numerous possible intermediate steps of their photodissociation. When only direct atomic H photolysis by breaking aromatic C-H bonds is considered, the value of N$_{\rm Ccr}$ could be as low as $\sim$32 (e.g., Table 4 
of  Castellanos et al.\ 2018a). It has been proven, however, that various processes implying H migration and/or H$_2$ formation may dominate for large PAHs (Castellanos et al.\ 2018a,b). In addition, it is predicted  that the formation of pentagonal rings is competitive with the loss of a hydrogen atom (Parneix et al.\ 2017). Other possible isomer structure changes in the carbon skeleton, such as ring opening or Stone-Wales defects, may also be considered. Although these effects are probably extremely difficult to model, it is possible that the actual photodissociation activation energy is lowered in this H or C structure reorganization, implying higher values for N$_{\rm Ccr}$. 
One argument supporting a high value of N$_{\rm Ccr}$ is the absence of any DIB compatible with the visible bands of C$_{42}$H$_{18}^+$ (HBC$^+$) measured by Steglich et al.\ (2011; see also Gredel et al.\ 2011), which might be explained by photodissociation. On the other hand, it appears to be unlikely that a PAH such as circumcoronene (C$_{54}$H$_{18}$) is dissociated in the diffuse ISM because its direct H  photodissociation rate is orders of magnitude lower than the IR emission (Castellanos et al.\ 2018a, Andrews et al.\ 2016). Therefore we consider a wide range of possible 
values of N$_{\rm Ccr}$, from $\sim$35 to $\sim$50.

\begin{table}[htbp]
      \caption[]{Typical orders of magnitude for rates of PAH chemical processes in the diffuse ISM$^a$.}
         \label{tab:lines}
            \begin{tabular}{ l c c  l    }
            \hline 
           \noalign{\smallskip}
 {\small Process} &   {\small rate$^{-1}$}  &{\tiny ref.}$^d$  &  {\small 
 Comments} \\
{\tiny   }  & {\small yr}   &   &  {\tiny }   \\
      \noalign{\smallskip}      
{\tiny   Photon absorption}   & {\small  0.1}  &  {\small  1} &  {\tiny 6-13.6~eV photons}   \\
{\tiny  Photoionization }   &{\small  1}   &  {\small  3} &  {\tiny }   \\
{\tiny  e$^-$ recombination}   & {\small  1}  & {\tiny 4} &  {\tiny }   \\
{\tiny  2$^{nd}$ photoionization }   & {\small  3}  & {\small  5} &  {\tiny }   \\
{\tiny   H accretion}$^{c}$   & {\small  20}  & {\small  2} &  {\tiny }   
\\
{\tiny  C$^+$ accretion}$^{bc}$ & {\small  5x10$^3$}  & {\small  8}  &  {\tiny }   \\
{\tiny   O accretion}$^{bc}$   & {\small  3x10$^4$}  & {\small  6} &  {\tiny }   \\
{\tiny   2-photon absorption}   & {\small 3x10$^7$}  &   &  {\tiny }   \\
{\tiny   C$_{\rm n}$H$_{\rm m}$ accretion}   &  {\small  $>$10$^7$}  & {\small  8} &  {\tiny  Total accretion rate of}\\
{\tiny   }  &    &   &  {\tiny small C-rich molecules}   \\
{\tiny  He$^+$ sputtering }   &  {\small  $\sim$10$^8$}  & {\small  7} &  
{\tiny }   \\
{\tiny  PAH merging}   &  {\small   $\sim$10$^8$}  &   &  {\tiny }\\
      \noalign{\smallskip}
            \noalign{\smallskip}
            \hline
           \end{tabular}
{\small \begin{list}{}{} 
\item[$^a$ Reaction rates are estimated for a typical PAH/PAH$^+$ of] 
\item[50 C atoms in a diffuse interstellar cloud with n$_{\rm H}$\,=\,50\,cm$^{-3}$,]
\item[T=50-100\,K, UV intensity G$_0$\,=\,3 Habing\,=\,1.6\,10$^{14}$\,erg\,cm$^{-3}$.]
\item[$^{b}$ Assumed abundances: O 3.2 10$^{-4}$; C$^+$ 1.4 10$^{-4}$.]
\item[$^{c}$ Global PAH accretion rates have been divided by 2, assum-]
\item[ing that neutral and cation PAH abundances are about half] 
\item[the total PAH abundance.]
\item[$^d$ References: 1 Tielens 2013; 2 Andrews et al.\  2016; 3 Leach]
\item[1996; 4 Montillaud et al.\ 2013; 5 Zhen et al.\ 2016; 6 Snow] \& \item[Bierbaum 2008; 7 Micelotta et al.\ 2010a; 8 This paper.]
\end{list}}
\end{table}

\subsection{Overall chemistry: Hierarchy of reaction rates}

Figure 1 and Table A.1 display typical orders of magnitude for the  main processes that the smallest PAHs, with about 50 C atoms, undergo in the diffuse ISM for averaged conditions: density n$_{\rm H}$\,=\,50\,cm$^{-3}$, temperature T\,=\,50-100\,K, and UV intensity G$_0$\,=\,3\,Habing\,=1.6\,10$^{14}$\,erg\,cm$^{-3}$. 
The reaction rates of the relevant processes range over nine orders of magnitude so that there is  a well-defined hierarchy.
Together with photo-processes, the main processes are listed below  (we do not discuss electron attachment and double ionization that are not very important in the diffuse ISM).

$\bullet$ {\bf H accretion.} 
Because of the overwhelming abundance of atomic hydrogen in the diffuse ISM, reactions between H and PAH$^+$ are by far the most frequent pure chemical reactions undergone by PAHs. As expected, they occur at a significant fraction of the Langevin rate, 
\begin{equation}
{\rm k_L = 2\pi\,e\,(\alpha_H/\mu )^{0.5} =  2\times 10^{-9} cm^3 s^{-1}}
,\end{equation}
where $\alpha_{\rm H}$\,=\,0.67\,10$^{-24}$cm$^3$ is the H polarizability and $\mu$\,=\,1 is the reduced atomic mass.  
However, we  did not consider the processes that form loosely bound overhydrogenated states or immediately eject an H$_2$ molecule (e.g., Castellanos et al.\ 2018b)  because they are irrelevant for fullerene formation. We address only the rehydrogenation of partially dehydrogenated PAHs. 

The actual rate of H association with large PAH cations is only about 7\% 
of this value, as measured in the laboratory by Demarais et al.\ (2014). We use  this value, 1.4$\times$10$^{-10}$\,cm$^3$s$^{-1}$, in Table A.1 and Fig.\ 1 for the most important H process,  rehydrogenation of PAHs through association of H with dehydrogenated PAH$^+$, as adopted, for instance, by  Montillaud et al.\ (2013) and Andrews et al.\ (2016). This process is implied in the discussion of PAH dehydrogenation of Appendix A.2.

$\bullet$ {\bf C$^+$ accretion.} 
The Langevin rate for the reaction C$^+$ $+$ PAH is about 6$\times$10$^{-9}$\,cm$^3$s$^{-1}$ with $\alpha$\,$\sim$\,1.5$\times$10$^{-24}$ N$_{\rm C}$ cm$^3$ (e.g.,\ Allamandola et al.\ 1989) and N$_{\rm C}$\,=\,50. 
For anthracene (C$_{14}$H$_{10}$), Canosa et al.\ (1995) measured that the total reaction rate (charge exchange plus C$^+$ accretion) is 3$\times$10$^{-9}$\,cm$^3$s$^{-1}$, which is close to the Langevin rate, while the branching ratio for C$^+$ accretion is 40\%. This ratio is expected to increase with the mass of the PAH. We may thus estimate that the rate of accretion of C$^+$ onto neutral PAHs in the diffuse ISM remains close to 2$\times$10$^{-9}$\,cm$^3$s$^{-1}$. This 
yields about 2$\times$10$^{-4}$\,yr$^{-1}$ for the global rate of carbon accretion onto PAHs in the standard diffuse ISM (Table A.1 and Fig.\ 1).

Many structures are possible for the addition of the carbon atom to the carbon network. The reaction  of C with naphtalene studied by Krasnokutski 
et al.\ (2017) shows that the formation of a seven-membered ring by insertion into a peripherical hexagon should be energetically favored with a total energy about -4\,eV for naphtalene. Such a structure with N$_{\rm C}$\,$\sim$\,40-50 should be unstable by interstellar photodissociation, but it could be stabilized by H addition.
However, as discussed below (Appendix A.4), the overall net C$^+$ growth rate, if any, cannot be higher than the O sputtering rate in order to sustain the PAH size  distribution. 
Because the C$^+$ accretion rate is significantly  higher than the O rate (Table A.1 and Fig.\ 1), most C$^+$ accretions therefore must eventually lead to the ejection of at least one carbon atom. Overall, it is difficult to predict whether the net result of C$^+$ accretions leads to growth or sputtering of the carbon skeleton, and this result could depend on the PAH structure and size.

$\bullet$ {\bf O sputtering.}
Atomic oxygen easily reacts with PAH cations. The results for a number of 
measured reactions O $+$ PAH$^+$ are summarized in Snow and Bierbaum (2008). Although CO is substantially produced in reactions with benzene and naphtalene cations, the direct result is only O accretion for coronene, with a rate 1.3$\times$10$^{-10}$\,cm$^3$s$^{-1}$. We may conservatively assume that the rate is at least comparable for heavier PAHs. This yields  about 3$\times$10$^{-5}$\,yr$^{-1}$ for the global rate of oxygen accretion onto PAHs in the standard diffuse ISM (Table A.1 and Fig.\ 1). 

Although there is no information on the initial product of this reaction, 
it is generally assumed that an epoxy binding on two adjacent C-C carbons 
is most probable. However, the epoxy binding energy is only $\sim$2.4\,eV 
(e.g., Li et al.\ 2006), so that the O-PAH association cannot survive the 
next UV photon absorption, unless the O atom forms a more stable binding. 
This is favored by its high mobility on the PAH surface, for which the hopping barrier is lower than about 1\,eV (Li et al.\ 2006). A ketone therefore appears as a possible outcome, especially for heavy PAHs. For low values of  N$_{\rm C}$, however, the ketone may be easily photodissociated with loss of CO, whose dissociation energy and barrier may be lower than 1.5\,eV and 4\,eV, respectively, as found by Chen et al.\ (2018) for bisanthenquinone. The net effect of reactions with O for the smallest stable PAHs with N$_{\rm C}$\,$\la$\,50-60 should therefore be carbon sputtering. We may assume that its rate is practically equal to the rate of O accretion, 
$\gamma_{\rm O}\,\sim$~0.7$\times$10$^{-4}$\,yr$^{-1}$, for relatively small PAHs with N$_{\rm C}$ close to N$_{\rm Ccr}$.

\subsection{Size distribution and dehydrogenation rate}
{\bf Size distribution.}
The information about the size distribution of interstellar PAHs mainly comes from the observed IR spectrum and the ratios of its main features, especially implying the 3.3\,$\mu$m band. However, despite the enormous amount of data accumulated on PAH IR spectra by successive space observatories, our understanding of the origin of this size distribution is still limited. As quoted by Draine et al.\ (2020), for example,
 the size distribution of interstellar PAHs  results from various  processes:  fragmentation of dust grains, chemisputtering, He$^+$ and e$^-$ sputtering in hot gas, growth by C$^+$ accretion, agglomeration with other PAHs, and accretion onto dust grains.... It is therefore impossible to derive a precise shape of the PAH size distribution because of our limited understanding of these processes. We therefore have to refer to empirical size distributions proposed on the basis of the IR feature ratios (e.g., D\'esert et al.\ (1990); Draine \& Li (2007); Maragkoudakis et al.\ (2020)).

These distributions and the corresponding PAH abundances as a function of N$_{\rm C}$, n$_{\rm PAH}$(N$_{\rm C}$)/n$_{\rm H}$, are well visualized in Fig.\ 5 of Pilleri et al.\ (2009), for example. We therefore  approximate the 
PAH abundance as n$_{\rm PAH}$(N$_{\rm C}$)/n$_{\rm H}$ = 10$^{-8}$ for 
N$_{\rm C}$\,$\le$\,50 and  10$^{-8}\times$(N$_{\rm C}$/50)$^{-1.67}$ for 
N$_{\rm C}$\,$>$\,50. The difference with more elaborate models is not significant compared to their uncertainty. This corresponds to X$_{\rm C}$\,$\sim$\,1.5$\times$10$^{-3}$ of total interstellar carbon for each value 
of  N$_{\rm C}$\,$\sim$\,50, that is,\ close to the dehydrogenation limit.
For higher values of N$_{\rm C}$, the distribution of \ X$_{\rm C}$(N$_{\rm C}$)  is expected to decrease mildly with N$_{\rm C}$, as (N$_{\rm C}$/50)$^{-0.67}$ for the size distribution we assumed above.

{\bf Overall PAH dehydrogenation rate.}
Maintaining such a high abundance of PAHs close to the photodissociation limit needs to compensate for the dehydrogenation sink by an overall net downsizing of PAHs because there is no down-top feeding from smaller fullerenes or other carbonaceous carbon clusters. When we assume that carbon loss cannot 
result from the only effect of single-photon photodissociation and H reactions, this net downsizing should be dominated by C$^+$ and O reactions because their rates are much faster than those of any others (Fig.\ 1).  Atomic oxygen sputtering is an obvious contribution to the net PAH downsizing. The overall effect of C$^+$ accretion is more uncertain because its net effect on the carbon skeleton may be either growth or sputtering.  C$^+$ growth, if any, must be slower than O sputtering to preserve a net downsizing, however. We note that this condition should also avoid the catastrophic depletion 
of gas carbon considered by Thaddeus (1995), for example. It is unlikely that  C$^+$ growth accidently almost compensates  for O sputtering. We may therefore consider that it is likely that some fraction of the O sputtering rate, 
for instance, 0.3, provides a conservative lower limit for the actual sputtering rate, that is,\ $\gamma_{\rm spPAH}$\,$\ga$\,0.3\,$\gamma_{\rm O}$\,=\,0.2$\times$10$^{-4}$\,yr$^{-1}$. Because the sputtering rate, $\gamma_{\rm spPAH}$, is also obviously the overall rate of PAH dehydrogenation, this is also a lower limit for the dehydrogenation rate.

\section{Computation of absorption spectra of fullerenes and nanotube segments}

\subsection{Fullerenes}
\begin{figure}[htbp]
         \begin{center}
\includegraphics[scale=0.62, angle=0]{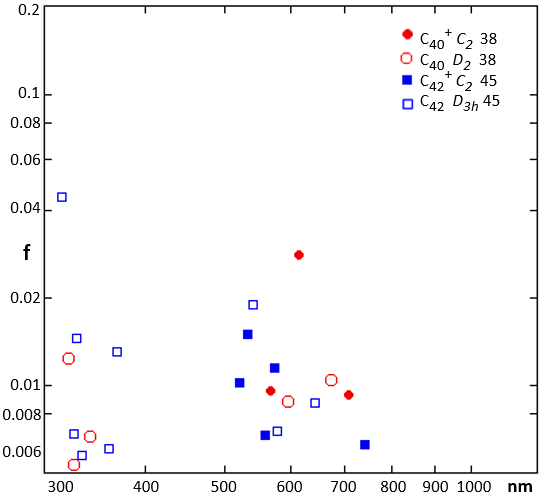}
 \caption{Computed absorption spectra (f-values) of neutral  fullerenes C$_{40}$ {\it D$_2$} 38 and C$_{42}$ {\it D$_{3h}$} 45 and cations C$_{40}^+$ {\it C$_2$} 38  and  C$_{42}^+$ {\it C$_{2}$} 45 (Table B.2) from $\sim$420\,nm to 1200\,nm.  Neutral isomers are displayed by open symbols and cations by full symbols.}
    \end{center}
 \end{figure}

\begin{figure}[htbp]
         \begin{center}
\includegraphics[scale=0.67, angle=0]{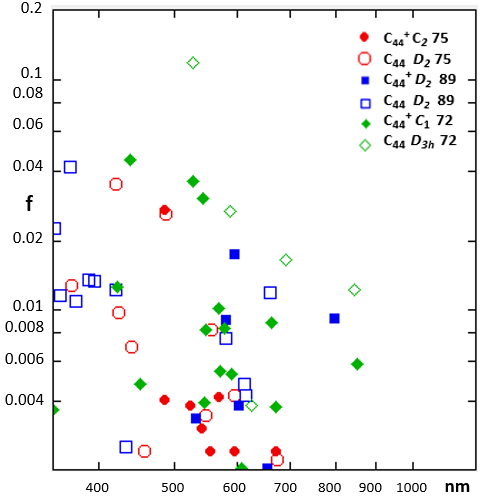}
 \caption{Computed absorption spectra (f-values) of three isomers of fullerene {\bf C$_{44}$} ({\it D$_2$} 75, {\it D$_2$} 89, and {\it D$_{3h}$} 72) and {\bf C$_{44}^+$} ({\it C$_2$} 75, {\it D$_2$} 89, and {\it C$_1$} 72) 
(Table B.2). Neutral isomers are displayed by open symbols and cations by 
full symbols.}
     \end{center}
 \end{figure}

\begin{figure*}[htbp]
         \begin{center}
\includegraphics[scale=0.73, angle=0]{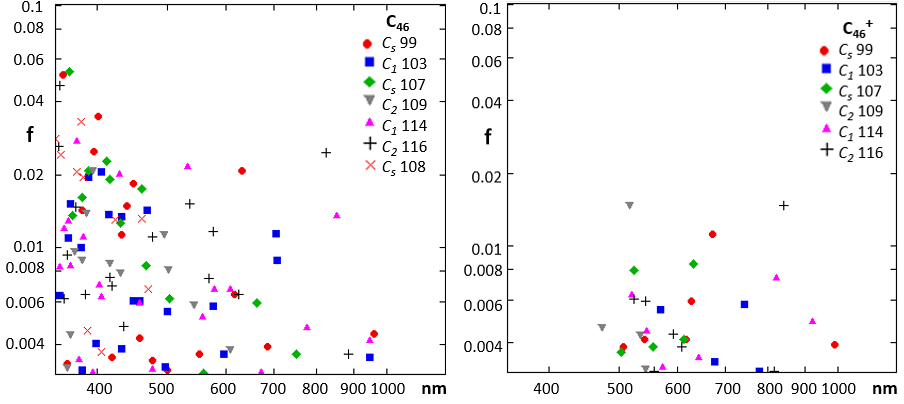}
 \caption{Computed absorption spectra (f-values) of neutral  fullerenes from 350\,nm to 1200\,nm, {\bf C$_{46}$}: {\it C$_s$} 99, {\it C$_1$} 103, 
{\it C$_s$} 107, {\it C$_s$} 108, {\it C$_2$} 109, {\it C$_1$} 114, and   {\it C$_2$} 116 {\it (left)}, and cations from $\sim$480\,nm to 1200\,nm, {\bf C$_{46}^+$}: {\it C$_s$} 99, {\it C$_1$} 103, {\it C$_s$} 107, {\it C$_2$} 109, {\it C$_1$} 114, and   {\it C$_2$} 116  {\it (right)} (Table B.2).}
     \end{center}
 \end{figure*}

\begin{figure*}[htbp]
         \begin{center}
\includegraphics[scale=0.73, angle=0]{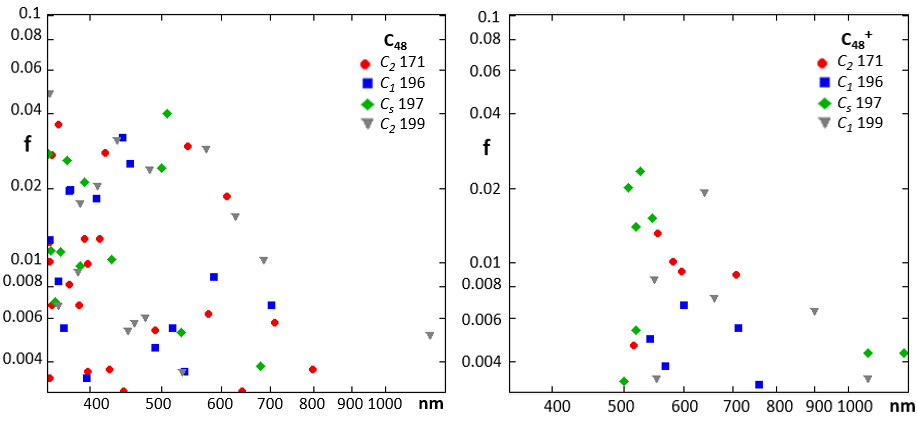}
 \caption{Computed absorption spectra (f-values)  of neutral fullerenes {\bf C$_{48}$}: {\it C$_2$} 171, {\it C$_1$} 196, {\it C$_s$} 197, and  {\it C$_2$} 199, from 350\,nm to 1200\,nm  {\it (left)}, and cations {\bf C$_{48}^+$}: {\it C$_2$} 171, {\it C$_1$} 196, {\it C$_s$} 197, and  {\it C$_1$} 
199, from $\sim$500\,nm to 1200\,nm {\it (right)} (Table B.2).}
     \end{center}
 \end{figure*}

\begin{table}[htbp]
      \caption[]{Energies (in eV) of first $^1$T$_{1u}$  states of C$_{60}$ buckminsterfullerene computed at various levels of theory.}
         \label{tab:lines}
            \begin{tabular}{ l l l }
            \hline
           \noalign{\smallskip}
Level &  1$^1$T$_{1u}$  &  2$^1$T$_{1u}$ \\
          \hline
& & \\
B3LYP/6-311$+$G**              &  3.44   &  3.88 \\
B3LYP/6-31G*                &  3.47      &  3.93 \\
CAM-B3LYP/6-311$+$G** & 4.28      &     4.65 \\
$\omega$B97X/6-311$+$G**               &  4.62   &  5.36 \\
$\omega$B97XD/6-311$+$G**        & 4.36       &  4.88 \\
PBE0/6-311$+$G**              &  3.62    &  4.01 \\
M062X/6-311$+$G**      &        4.26     &  4.58 \\
CASPT2$^a$                 &   3.09      &  3.19 \\ 
experiment$^b$        &  3.08    &  3.30 \\
      \noalign{\smallskip}
            \hline
           \end{tabular}
{\small \begin{list}{}{} 
\item[$^a$ Using an atomic natural orbital basis of $3s2p1d$ quality,] 
\item[see Anderson (2020).] 
\item[$^b$ As given in Anderson (2020).] 
\end{list}}
\end{table}

As quoted in Section 5.2, in order to estimate the possibility of measuring the abundances of  interstellar fullerenes through their optical absorption, 
we performed B3LYP/6-31G* calculations of the visible spectrum for 19 neutral fullerene isomers (and their cations except for one), with even values of N$_{\rm C}$ between 40 and 50 (Table B.2).  

For each of the fullerenes C$_{40}$ to C$_{50}$, the structures of those isomers that are within 10 kcal mol$^{-1}$ in energy with respect to the most stable isomer as determined previously by Sun et al.\ (2005, see Table B.2) were optimized using the hybrid-density functional B3LYP (Becke 1993; Lee et al.\ 1988) in conjunction with the 6-31G* basis set implemented in Gaussian 16 (Frisch et al.\ 2016). For comparison, the structure of the I$_h$ isomer of C$_{60}$, buckminsterfullerene was also optimized at 
this level of theory.

        The energy of excited states and oscillator strengths were computed using the optimized structures and time-dependent density functional theory (TD-DFT). We employed various functionals [B3LYP, CAM-B3LYP (Yanai et al.\ 2004), $\omega$B97X (Chai et al.\ 2008a), $\omega$B97XD (Chai et al.\ 
2008b),  PBE0 (Adamo \& Barone 1999; Ernzerhof \& Scuseria 1999),  and M062X (Zhao \& Truhlar 2008)] in conjunction with the 6-311$+$G** basis set for computing the electric dipole allowed transitions from the ground state to the two lowest lying $^1$T$_{1u}$ states of buckminsterfullerene to 
compare with experimental and sophisticated CASPT2 (Anderson 2020) data (Table B.1). 

We find that the deviation of the B3LYP values from experiment is the lowest of the functionals we investigated (0.4\,eV and 0.6\,eV for 1$^1$T$_{1u}$  and  2$^1$T$_{1u}$, respectively). We also note that the energies hardly change when the much more economical 6-31G* basis set is employed. We thus chose the B3LYP/6-31G* level of theory to compute the excited 
state energies of the fullerenes for the semiquantitative purpose of this study.

        The structures of the fullerene radical cations were optimized at the B3LYP/6-31G* level of theory starting from the structure of the neutrals without symmetry constraint.  
Vibrational analysis confirmed that they represent local minima of the potential energy surface. The point groups of the derived structures were directly identified, and they are  reported in Table B.2.  For C$_{44}^+$ and C$_{50}^+$, these point groups agree with those given by Candian et al. (2019).
Subsequently, the excited state energies were computed with the TD-DFT approach employing the B3LYP functional and the 6-31G* basis set.

        The structures of a few single-wall carbon nanotube segments were similarly optimized at the B3LYP/6-31G* level of theory for the neutrals and cations, and the excited state energies were subsequently computed at the B3LYP/6-31G* level of theory. The results are presented in Appendix B.2.

\begin{table}[htbp]
      \caption[]{Fullerene isomers included in the calculations$^a$.}
         \label{tab:lines}
            \begin{tabular}{ l r l l  l l l l}
            \hline
           \noalign{\smallskip}
 {\small N$_{\rm C}$}& n$_{\rm iso}^b$&sym$^c$&{\small $\Delta{\rm E}^d$}&{\small N$_{\rm C}$}& n$_{\rm iso}^b$&sym$^c$&{\small $\Delta{\rm E}^d$} 
\\
      \noalign{\smallskip}
C$_{40}$&38&$D_2 C_2$&$0.0$&C$_{42}$&45&$D_3 C_2$&$0.0$\\
&&&&&&&\\
C$_{44}$&75&$D_2 C_2$&$0.0$&C$_{46}$&109&$C_2 C_2$&$0.0$\\
&89&$D_2 D_2$&$0.7$&&108$^e$&$C_s$&$2.0$\\
&72&$D_{3h} C_1$&$7.8$&&103&$C_1 C_1$&$5.9$\\
&&&&&107&$C_s C_s$&$6.0$\\
&&&&&114&$C_1 C_1$&$6.2$\\
&&&&&116&$C_2 C_2$&$7.6$\\
&&&&&99&$C_s C_s$&$7.4$\\
&&&&&&&\\
C$_{48}$&171&$C_2 C_2$&$0.0$&C$_{50}$&270&$D_3 C_2$&$0.0$\\
&199&$C_2 C_1$&$2.6$&&271&$D_{5h}C_{5h}$&$5.6$\\
&196&$C_1 C_1$&$2.9$&&266&$C_s C_s$&$8.2$\\
&197&$C_s C_s$&$5.2$&&&&\\ 
      \noalign{\smallskip}
            \hline
           \end{tabular}
{\small \begin{list}{}{} 
\item[$^a$ The spectra of all listed neutral and cation isomers have] 
\item[been calculated, except for C$_{44}^+$ 108. For each value of N$_{\rm C}$,] 
\item[they are listed in order of increasing values of formation]
\item[energies given by Sun et al.\ (2005).]
\item[$^b$ Isomer number from Fowler \& Manolopoulos (1995).]
\item[$^c$ Point group symmetry: 1) neutral; 2) radical cation.]  
\item[$^d$ Relative electronic energy (kcal/mol).] 
\item[$^e$ The cation spectrum was not calculated.]
\end{list}}
\end{table}

All calculated spectra of C$_{40}$ to C$_{50}$ fullerenes display at least one and generally several significant bands between 400 and 800 nm, with f-values higher than 10$^{-2}$ (Figs.\ 2 and B.1 to B.4). There is a trend for finding stronger bands for the most stable fullerenes, C$_{50}$ and C$_{44}$, and with increasing values of N$_{\rm C}$. C$_{50}$ isomers have several visible bands with f $\sim$ (5-8)$\times$10$^{-2}$. Most of these relatively strong bands are concentrated at $\lambda$ $\la$ 550 nm, 
but a few are found at $\sim$550-600 nm (Fig.\ 2). The visible bands of the isomers of the other fullerenes display a similar wavelength distribution, but with significantly lower f-values, by a factor $\sim$2 for C$_{44}$ and C$_{48}$, $\sim$3 for C$_{46}$, and  $\sim$6 for C$_{40}$ and  C$_{42}$ (see Figures B.1 to B.4). 

Our calculations  generally confirm some similarity of the spectra of cation and neutral forms of the same molecule, as is known for PAHs (e.g., Malloci et al.\ {\b f2004).}
While the f-value of each the C$_{60}^+$ bands at 9577/9632\,\AA\  is $\sim$3-2$\times$10$^{-2}$ (Campbell \& Maier 2017), similar f-values are found for two bands of C$_{50}^+$ $D_{5h}$ at $\lambda$ $>$ 600 nm, and about half the computed cation spectra include a band at $\lambda$ $>$ 600 nm with f\,$\ga$\,10$^{-2}$.

\begin{figure}[htbp]
         \begin{center}
\includegraphics[scale=0.75, angle=0]{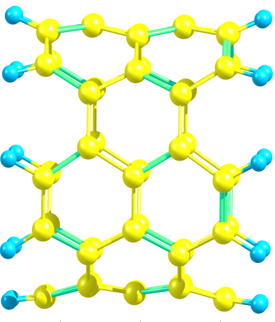}
 \caption{Example of nanotube segment: Armchair (5,5) C$_{50}$H$_{20}$}
     \end{center}
 \end{figure}

\subsection{Nanotube segments}

\begin{figure}[htbp]
         \begin{center}
\includegraphics[scale=0.62, angle=0]{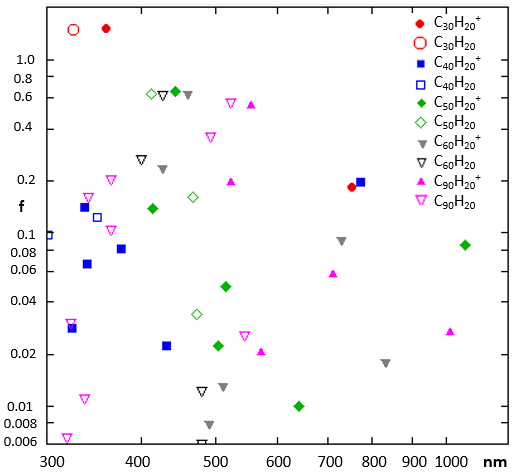}
 \caption{Computed absorption spectra (f-values) of some (5,5) armchair nanotube segments, C$_{\rm n}$H$_{20}$, with n\,=\,30, 40, 50, 60, and  90. Full symbols show cations, and open symbols show neutrals.}
     \end{center}
 \end{figure}

\begin{figure}[htbp]
         \begin{center}
\includegraphics[scale=0.65, angle=0]{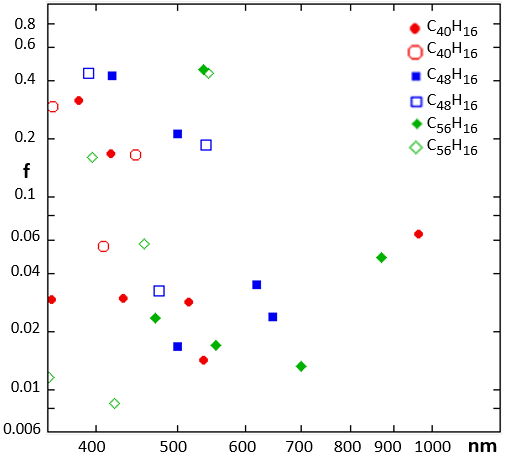}
 \caption{Computed absorption spectra (f-values) of some (4,4) armchair nanotube segments, C$_{\rm n}$H$_{16}$, with n\,=\,40, 48, and 56. Full symbols show cations, and open symbols show neutrals.}
     \end{center}
 \end{figure}

\begin{figure*}[htbp]
         \begin{center}
\includegraphics[scale=0.8, angle=0]{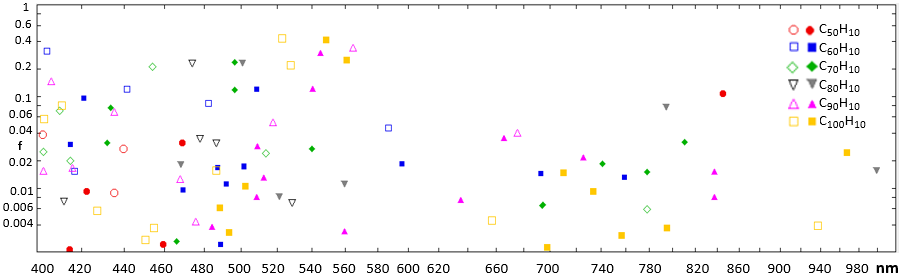}
 \caption{Computed absorption spectra (f-values) of some semicapped (5,5) armchair nanotube segments, C$_{\rm n}$H$_{10}$, with n\,=\,50, 60, 70, 80, 90, and  100. Full symbols show cations, and open symbols show neutrals. The lower $\lambda$ limit of the computations is 400\,nm for all neutrals. For cations, it varies from $\sim$400\,nm for C$_{50}$H$_{10}^+$ to $\sim$480\,nm 
for C$_{100}$H$_{10}^+$.}
     \end{center}
 \end{figure*}

Nanotube segments (e.g., Fig.\ B.5) present similarities with symmetrical PAHs and fullerenes. Their strong visible bands justify their consideration as DIB carriers (Zhou et al.\ 2006).  As their formation appears to be possible from partially dehydrogenated PAHs (Section 3.2; Chen et al.\ 2020), we calculated the optical absorption spectra of a few of the shortest 
neutral and cation (4,4) and (5,5) armchair tube-segments with the same method as fullerenes (Appendix B.1).
 Figs.\ B.6 and B.7 display the spectra of (4,4) C$_{\rm n}$H$_{16}$, with n\,=\,40, 48, and  56 and  (5,5) C$_{\rm n}$H$_{20}$, with n\,=\,30, 40, 50, 60, and  90. 
 All these neutral and cation tube-segments with N$_{\rm C}\,>$\,40 have a strong band in the range $\sim$400-800\,nm with f\,$\sim$\,0.5. This confirms the results of Zhou et al.\ (2004, 2006) for (5,5) armchair tube-segments, but with significantly lower f-values. There are many additional weaker bands, including cation bands at $\lambda$\,$\ga$\,550\,nm with f\,$\ga$\,0.02.   
The size distribution of these nanotube-segments and the resulting interstellar absorption spectrum is  expected to depend not only on their formation, but also on their subsequent chemistry and possible growth by C$^+$ accretion. It is not excluded that they might carry some  DIBs if they formed in the diffuse ISM.

We extended the calculations to a few neutral and cation semicapped  armchair (5,5) nanotube segments, C$_{\rm n}$H$_{10}$, with n\,=\,50, 60, 70, 80, 90, and  100, including C$_{50}$H$_{10}$, which was synthesized by Scott et al.\ (2011). 
Their visible absorption spectra (Fig.\ B.8) have similar general characteristics to open nanotube segments (Figs.\ B.6 and B.7), with a trend to slightly weaker features. The computed spectrum of neutral C$_{50}$H$_{10}$  appears to be compatible with that measured by Scott et al.\ (2011) in CH$_2$Cl$_2$.

\subsection{Summary of fullerene chemistry}
The chemistry of interstellar fullerenes was addressed in Omont (2016), but mostly in view of  C$_{60}$. The main conclusions are that many atoms or ions, including H, C$^+$, and O, probably accrete onto the cage, either ionized or neutral, with a rate comparable to the Langevin rate, similarly to PAHs (Appendix A.3). Low binding energies and the high C$_{60}$ stability prevent the formation of stable compounds, however. Most accreted species are just quickly lost through photodissociation. The only exception might be Fe$^+$, while it is not completely excluded that atomic oxygen 
might lead to sputtering through CO ejection. 

Smaller fullerenes should have a similar uncertain chemistry, but their lower stability  might increase the possibility of sputtering, especially following reactions with atomic oxygen. The case of C$^+$ was quoted in Section 4.1. Its accretion rate could be similar to PAHs and might play an important role in cage evolution. However, its accretion is thought to remain transitory, before its ejection by the next UV photon with E\,$\ga$\,10\,eV. Cage sputtering through loss of C$_3$ is even possible for 
low values of N$_{\rm C}$, but this is uncertain. Cage growth by incorporation of a C$_2$ unit after the accretion of a second C$^+$ therefore appears to be inefficient. Replenishment in small PAHs from other types of C$_{\rm n}$ clusters with n\,$<$\,N$_{\rm Ccr}$ also appears to be unlikely because graphene flakes are probably unstable and tubular structures cannot easily return to the planar carbon skeleton of PAHs.
Hydrogenated fullerenes (fulleranes) are probably more unstable than with C$_{60}$, and the formation of interstellar endohedral compounds appears to be at least as difficult as for C$_{60}$. 

In conclusion, cage decay of these fullerenes could be  as rapid as for PAHs for N$_{\rm C}$\,$\sim$\,50, but probably not faster. Direct photodissociation decay through C$_2$ loss might be considered for  N$_{\rm C}$\,$<$\,40.

\end{document}